\def\apjl{Astrophys. J. Lett.}
\def\apjs{Astrophys.J.Suppl.}

\def\aap{Astron. Astrophys.}

\newcommand{\ra}{\;\raise1.0pt\hbox{$'$}\hskip-6pt\partial\;}
\newcommand{\lo}{\;\overline{\raise1.0pt\hbox{$'$}\hskip-6pt\partial}\;}

\documentclass[traditabstract]{aa} 
\usepackage{graphicx}
\usepackage{txfonts}
\usepackage{natbib}

\begin{document}
\title{How to make a clean separation between CMB E and B modes with proper foreground masking} 
\titlerunning{How to achieve a clean CMB E/B separation with proper foreground masking}
\subtitle{}
\author{Jaiseung Kim\thanks{jkim@nbi.dk}}
\institute{Niels Bohr Institute \& Discovery Center, Blegdamsvej 17, DK-2100 Copenhagen, Denmark}

\abstract
{
We investigate the E/B decomposition of CMB polarization on a masked sky. 
In real space, operators of E and B mode decomposition involve only differentials of CMB polarization.
We may, therefore in principle, perform a clean E/B decomposition from incomplete sky data. 
Since it is impractical to apply second derivatives to observation data, we usually rely on spherical harmonic transformation and inverse transformation, instead of using real-space operators. In spherical harmonic representation, jump discontinuities in a cut sky produces Gibbs phenomenon, unless a spherical harmonic expansion is made up to an infinitely high multipole. By smoothing a foreground mask, we may suppress the Gibbs phenomenon effectively in a similar manner to apodization of a foreground mask discussed in other works. However, we incur foreground contamination by smoothing a foreground mask, because zero-value pixels in the original mask may be rendered non-zero by the smoothing process. In this work, we investigate an optimal foreground mask, which ensures proper foreground masking and suppresses Gibbs phenomenon.
We apply our method to a simulated map of the pixel resolution comparable to the Planck satellite. The simulation shows that the leakage power is lower than unlensed CMB B mode power spectrum of tensor-to-scalar ratio $r\sim 1\times10^{-7}$. 
We compare the result with that of the original mask.
We find that the leakage power is reduced by a factor of $10^{6} \sim 10^{9}$ at the cost of a sky fraction $0.07$, and that the enhancement is highest at lowest multipoles. We confirm that all the zero-value pixels in the original mask remain zero in our mask.
The application of this method to the Planck data will improve the detectability of primordial tensor perturbation.}

\maketitle

\section{Introduction}
Over the past few years, CMB polarization has been measured by several experiments, and is being measured by the Planck surveyor \citep{DASI:data,DASI:instrument,DASI:I,DASI:II,DASI:III,DASI:3yr,QUaD1,QUaD2,QUaD:instrument,QUaD_improved, Planck_bluebook}.
Using the general properties of symmetric trace-free tensors, we may decompose CMB polarization into the gradient-like E mode and the curl-like B mode \citep{Kamionkowski:Flm,Seljak-Zaldarriaga:Polarization}.
In the standard model, B mode polarization is not produced by scalar perturbation, but solely by tensor perturbation. 
Therefore, the measurement of the B mode polarization makes it possible to probe the universe on the energy scale during the inflationary period \citep{Kamionkowski:Flm,Seljak-Zaldarriaga:Polarization,Modern_Cosmology,Inflation,Foundations_Cosmology}. 
We can also obtain information about the dark matter distribution from weak lensing imprints on B mode polarization \citep{CMB_Lensing:full-sky,B_lensing,CMB_Lensing}.
In most inflationary models, the tensor-to-scalar ratio $r$ is much smaller than one, and the current upper bound $r<0.36$ is imposed by the WMAP 7 year data at a $95\%$ confidence level \citep{WMAP7:powerspectra,WMAP7:Cosmology}. 
Therefore, it is quite a challenging and ambitious goal to measure CMB B mode polarization.

Besides contamination inherent to instruments such as noise and instrumental polarization, several complications limit the detectability of tensor perturbation.
In particular, there is contamination from Galactic and extragalactic foregrounds \citep{B_mode_limit_foreground,tensor_limit_fg,B_limit_inflation}.
By spectral matching component separation or template fitting, we may clean foreground contamination \citep{TS_ratio_fg,B_lowl_fg,B_forecast_large}.
However, we need to mask out some regions that cannot be cleaned reliably. For the analysis of the WMAP data, the WMAP team cleaned diffuse foreground by template fitting, and masked out 20$\sim$30\% of a whole sky. 
Incomplete sky coverage, due to foreground masking, leads to E/B mixing\citep{Bunn:EB-Separation,EB_incomplete_sky}. 
Therefore, there have been various efforts to understand and reduce E/B mixing associated with incomplete sky coverage \citep{Bunn:EB-Separation,EB_incomplete_sky,EB_harmonic,Smith:pseudo_EB,Kim:optimization,Kim:measuring_a2lm,EB_pixel,EBpixel_Zhao,EBpixel_Bunn}.
Previously, we investigated E/B decomposition in pixel space, and showed that E/B mixing is localized mostly around the boundary of foreground cuts \citep{EB_pixel}. 

In real space, E and B mode decomposition operators involve only the differential of CMB polarization.
Therefore, in principle, we may succeed in performing a clean E and B decomposition in real space. 
Since it is impractical to apply second derivatives to observation data, we usually perform a spherical harmonic transformation and an inverse transformation to achieve E/B decomposition, instead of applying real-space operators.
In spherical harmonic representation, jump discontinuities in a cut sky produces Gibbs phenomenon, unless spherical harmonic expansion consists of infinitely high multipoles. However, we may suppress the Gibbs phenomenon effectively by smoothing a foreground mask, which is similar to the apodization of a foreground mask discussed in other works, though developed in a slightly different context \citep{Smith:pseudo_EB,edge_taper,EB_pixel,EBpixel_Zhao}.
By smoothing a foreground mask, we incur foreground contamination, because zero-value pixels in the original mask may be rendered non-zero by the smoothing process.
Therefore, we investigate an optimal foreground mask, which ensures proper foreground masking without any unnecessary loss of sky fraction and suppresses Gibbs phenomenon at the same time.
We apply our method to a simulated map of the pixel resolution comparable to the Planck satellite. The simulation shows that leakage power in unmasked pixels is comparable to or smaller than the unlensed CMB B mode power spectrum of the tensor-to-scalar ratio $r\sim 1\times10^{-7}$. 
Compared with the result of the original mask, we have reduced the leakage power by a factor of $10^{6} \sim 10^{9}$ at the cost of only a sky fraction $0.07$.
We find that the enhancement is highest at the lowest multipoles and that all the zero-value pixels of the original mask remain zero in our mask. 

The outline of this paper is as follows. 
In Sec. \ref{Stokes}, we discuss CMB polarization and E/B decomposition.
In Sec. \ref{incomplete_sky}, we discuss E/B decomposition of a masked sky and the Gibbs phenomenon. 
In Sec. \ref{discrepancy}, we investigate the degree of the Gibbs phenomenon in a cut-sky CMB map. 
In Sec. \ref{smoothing}, we present a rigorous discussion of an optimal foreground mask. 
In Sec. \ref{simulation}, we apply our method to simulated data and present our results.
In Section \ref{Discussion}, we summarize our investigation. 

\section{STOKES PARAMETERS}
\label{Stokes}
The state of polarization is described by Stokes parameters \citep{Kraus:Radio_Astronomy,Tools_Radio_Astronomy}.
Since Thompson scattering does not generate circular polarization, Stokes parameter Q and U are sufficient to describe CMB polarization \citep{Modern_Cosmology}. 
For the rotation of an angle $\psi$ on the plane perpendicular to direction $\mathbf n$, Stokes parameter $Q$ and $U$ have the following spin $\pm$2 properties  \citep{Seljak-Zaldarriaga:Polarization,Zaldarriaga:Polarization_Exp}.
\begin{eqnarray}
\label{Q'U'} (Q\pm i U)'(\mathbf n)=e^{\mp 2i\psi}(Q\pm i U)(\mathbf n).
\end{eqnarray}
Using its spin properties, we may decompose all-sky Stokes parameters in terms of spin $\pm2$ spherical harmonics \citep{Seljak-Zaldarriaga:Polarization}:
\begin{eqnarray}
Q(\mathbf n)\pm i U(\mathbf n)&=&\sum_{l,m} a_{\pm2,lm}\;{}_{\pm2}Y_{lm}(\mathbf n),\label{Q_lm+iU_lm} 
\end{eqnarray}
where the decomposition coefficients $a_{\pm2,lm}$ are obtained by:
\begin{eqnarray}
a_{\pm2,lm}=\int \left[Q(\mathbf n)\pm i U(\mathbf n)\right]\,{}_{\pm2}Y^*_{lm}(\mathbf n)\,\mathrm d \mathbf n.\label{a2lm}
\end{eqnarray}
Though the quantity shown in Eq. \ref{Q_lm+iU_lm} has a direct association with physical observables (i.e. Stokes parameters),
it is desirable to derive rotationally invariant scalar quantities.
To derive spin-zero quantities, we may apply the spin raising and lowering operators given by \citep{Seljak-Zaldarriaga:Polarization}.
\begin{eqnarray}
\ra {}_s f(\theta,\phi)&=&-\sin^s \theta\left[\frac{\partial}{\partial \theta} +i \csc\theta \frac{\partial}{\partial \phi}\right] \sin^{-s}\theta\;{}_s f(\theta,\phi),\label{ra}\\
\lo {}_s f(\theta,\phi)&=&-\sin^{-s} \theta\left[\frac{\partial}{\partial \theta} -i \csc\theta \frac{\partial}{\partial \phi}\right] \sin^{s}\theta\;{}_s f(\theta,\phi)\label{lo},
\end{eqnarray}
where ${}_s f(\theta,\phi)$ is an arbitrary spin $s$ function. 
Applying these operators to $Q(\mathbf n)\pm i U(\mathbf n)$ sequentially, we may construct the following scalar quantities \citep{Kamionkowski:Flm,Seljak-Zaldarriaga:Polarization}:
\begin{eqnarray}
\lefteqn{E(\mathbf n)=\nonumber}\\
&&-\frac{1}{2}[\lo^2 (Q(\mathbf n) + i U(\mathbf n))+\ra^2 (Q(\mathbf n)- i U(\mathbf n))],\label{E_map}\\
\lefteqn{B(\mathbf n)=\nonumber}\\
&&\frac{i}{2}[\lo^2 (Q(\mathbf n) + i U(\mathbf n))-\ra^2 (Q(\mathbf n)- i U(\mathbf n))]\label{B_map}.
\end{eqnarray}
These two scalar quantities are often termed `E' and `B' mode, and associated, respectively, with gradient-like and curl-like components of the CMB polarization pattern.
Besides rotational invariance, the construction of scalar B quantities increases the detectability of the primordial tensor perturbation, because
primordial scalar perturbation makes a null contribution to the B mode \citep{Seljak-Zaldarriaga:Polarization}.

Applying the spin operators to spin-weighted spherical harmonics, we find the properties \citep{Seljak-Zaldarriaga:Polarization}.
\begin{eqnarray}
\ra {}_{s}Y_{lm}(\mathbf n)&=&\sqrt{(l-s)(l+s+1)}\;\;{}_{s+1}Y_{lm}(\mathbf n), \label{ra_sYlm}\\
\lo {}_{s}Y_{lm}(\mathbf n)&=&-\sqrt{(l+s)(l-s+1)}\;\;{}_{s-1}Y_{lm}(\mathbf n).\label{lo_sYlm}
\end{eqnarray}
Using Eqs. \ref{E_map}, \ref{B_map}, \ref{ra_sYlm}, and \ref{lo_sYlm}, we find that the `E' and `B' mode quantities are equivalently given by 
\citep{Seljak-Zaldarriaga:Polarization}.
\begin{eqnarray}
E(\mathbf n)&=&\sum_{lm} \sqrt{\frac{(l+2)!}{(l-2)!}}\,a_{E,lm}\,Y_{lm}(\mathbf n),\label{Elm}\\
B(\mathbf n)&=&\sum_{lm} \sqrt{\frac{(l+2)!}{(l-2)!}}\,a_{B,lm}\,Y_{lm}(\mathbf n),\label{Blm}
\end{eqnarray}
where
\begin{eqnarray}
a_{E,lm}&=&-(a_{2,lm}+ a_{-2,lm})/2,\label{alm_E}\\
a_{B,lm}&=&i(a_{2,lm} - a_{-2,lm})/2.\label{alm_B}
\end{eqnarray}
In real-world observations, angular-scale anisotropy much smaller than the FWHM (full width at half maximum) of physical beams are sufficiently suppressed. 
For instance, the window function of the FWHM $15^\prime$ at the multipole $l=4000$ has a value $\sim 10^{-24}$.
Therefore, summations in Eqs. \ref{Q_lm+iU_lm}, \ref{Elm}, and \ref{Blm} may be truncated to a finite multipole with good accuracy.
Since it is impractical to take second derivatives of observation data, we resort to Eqs. \ref{Elm}, \ref{Blm}, \ref{alm_E}, and \ref{alm_B}
to perform the E/B decomposition \citep{Kamionkowski:Flm,Foundations_Cosmology}.

\begin{figure}[htb!]
\includegraphics[scale=.43]{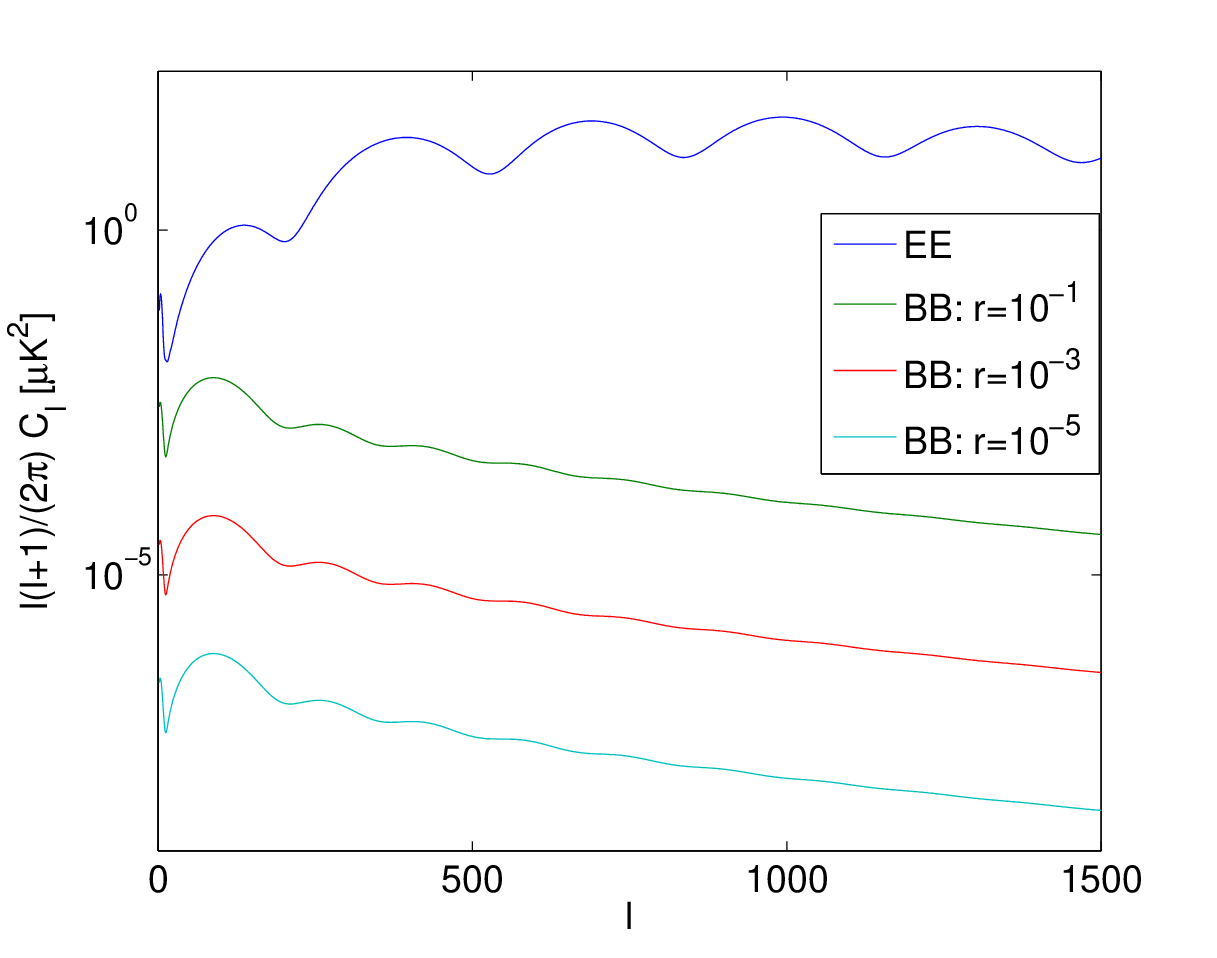}
\caption{The power spectrum of E and B: B mode power spectrum is plotted for various tensor-to-scalar ratio $r$.} 
\label{Cl}
\end{figure}
For a Gaussian seed fluctuation model, the decomposition coefficients of E and B mode satisfy the statistical properties as follows.
\begin{eqnarray} 
\langle a^*_{E,lm} a_{E,l'm'} \rangle &=& C^{EE}_l\,\delta_{ll'}\delta_{mm'},\\
\langle a^*_{B,lm} a_{B,l'm'} \rangle &=& C^{BB}_l\,\delta_{ll'}\delta_{mm'},
\end{eqnarray}
where $\langle\ldots\rangle$ denotes an ensemble average.
In Fig. \ref{Cl}, we show the unlensed $C^{EE}_l$ and  $C^{BB}_l$ values of the WMAP concordance $\Lambda$CDM model for various tensor-to-scalar ratio $r$.
As noted from Fig. \ref{Cl} and the WMAP7 upper bounds on $r<0.36$, we expect the CMB B mode polarization to be much smaller than that of the E mode.

\section{E and B mode decomposition of a masked sky}
\label{incomplete_sky}
Several astrophysical emission sources (the ``foreground'') exist between the last scattering surface and our vantage point. Contamination from the foregrounds degrades the cosmological information that is attainable from CMB data. 
Therefore, the WMAP team have reduced diffuse foregrounds by template-fitting, and blocked some regions by applying a foreground mask.
That is, reliable measurement of CMB polarization is not available over the whole sky, but only a masked sky.
We consider here E and B mode decomposition from a masked sky.
\begin{eqnarray}
\lefteqn{\tilde E(\mathbf n)=\nonumber}\\
&&-\frac{1}{2}[\lo^2 (\tilde Q(\mathbf n) + i\tilde U(\mathbf n))+\ra^2 (\tilde Q(\mathbf n)- i\tilde U(\mathbf n))],\label{E_same}\\
\lefteqn{\tilde B(\mathbf n)=\nonumber}\\
&&\frac{i}{2}[\lo^2 (\tilde Q(\mathbf n) + i\tilde U(\mathbf n))-\ra^2 (\tilde Q(\mathbf n)- i\tilde U(\mathbf n))]\label{B_same},
\end{eqnarray}
where
\begin{eqnarray}
\tilde Q(\mathbf n) \pm i \tilde U(\mathbf n)= W(\mathbf n)(Q(\mathbf n) \pm i U(\mathbf n)).\label{iQU}
\end{eqnarray}
If the sky direction $\mathbf n$ belongs to the bounded region $\mathcal R$ of foreground cuts, a foreground mask $W(\mathbf n)$ is set to zero, or otherwise one.
As shown in Eqs. \ref{ra} and \ref{lo}, a spin raising and lowering operator are, in fact, second derivatives weighted by trigonometric functions.
Therefore, a spin raising and lowering operator requires information within the infinitesimal vicinity of $\mathbf n$. 
Noting this, we may show that
\begin{eqnarray}
\tilde E(\mathbf n)&=&-\frac{1}{2}[\lo^2 W(\mathbf n)(Q(\mathbf n) + i U(\mathbf n))+\ra^2 (Q(\mathbf n)- i U(\mathbf n))],\\
&=&-\frac{1}{2}[\lo^2 (Q(\mathbf n) + i U(\mathbf n))+\ra^2 (Q(\mathbf n)- i U(\mathbf n))], \label{iE}
\end{eqnarray}
\begin{eqnarray}
\tilde B(\mathbf n)&=&\frac{i}{2}[\lo^2 W(\mathbf n)(Q(\mathbf n) + i U(\mathbf n))-\ra^2 (Q(\mathbf n)- i U(\mathbf n))],\\
&=&\frac{i}{2}[\lo^2 (Q(\mathbf n) + i U(\mathbf n))-\ra^2 (Q(\mathbf n)- i U(\mathbf n))], \label{iB}
\end{eqnarray}
where $\mathbf n$ belongs to a region outside the foreground cuts and their boundaries.
In other words, the E and B mode maps constructed from a masked sky are identical to those of a whole sky, as far as the concerned sky direction does not belong to the foreground cuts or their boundaries:
\begin{eqnarray}
\tilde E(\mathbf n)&=&E(\mathbf n),\\
\tilde B(\mathbf n)&=&B(\mathbf n),
\end{eqnarray}
where $\mathbf n\notin \mathcal R$.

Whether it is a whole-sky coverage or not, it is impractical to takes second derivatives of observation data.
Therefore, we perform a  E/B decomposition using Eqs. \ref{Elm}, \ref{Blm}, \ref{alm_E}, and \ref{alm_B} \citep{Kamionkowski:Flm,Foundations_Cosmology}.
We may equivalently construct E/B decomposed maps by
 \begin{eqnarray}
\tilde E(\mathbf n)&=&\sum_{lm} \sqrt{\frac{(l+2)!}{(l-2)!}}\,\tilde a_{E,lm}\,Y_{lm}(\mathbf n),\label{iElm}\\
\tilde B(\mathbf n)&=&\sum_{lm} \sqrt{\frac{(l+2)!}{(l-2)!}}\,\tilde a_{B,lm}\,Y_{lm}(\mathbf n),\label{iBlm}
\end{eqnarray}
where
\begin{eqnarray}
\tilde a_{E,lm}&=&-(\tilde a_{2,lm}+\tilde a_{-2,lm})/2,\label{ialm_E}\\
\tilde a_{B,lm}&=&i(\tilde a_{2,lm}-\tilde a_{-2,lm})/2,\label{ialm_B}
\end{eqnarray}
and
\begin{eqnarray}
\tilde a_{\pm2,lm}=\int \left[\tilde Q(\mathbf n)\pm i \tilde U(\mathbf n)\right]\,{}_{\pm2}Y^*_{lm}(\mathbf n)\,\mathrm d \mathbf n.\label{ia2lm}
\end{eqnarray}
As pointed out above, $\tilde E(\mathbf n)$ and $\tilde B(\mathbf n)$ are identical to true E and B mode maps for $\mathbf n\notin \mathcal R$. 
Therefore, E and B mode maps constructed by Eqs. \ref{iElm}, \ref{iBlm}, \ref{ialm_E}, and \ref{ialm_B} are identical to true E and B maps (i.e. those of a whole sky map) for $\mathbf n\notin \mathcal R$.
 \begin{eqnarray}
E(\mathbf n)&=&\sum_{lm} \sqrt{\frac{(l+2)!}{(l-2)!}}\,\tilde a_{E,lm}\,Y_{lm}(\mathbf n),\label{E_iElm}\\
B(\mathbf n)&=&\sum_{lm} \sqrt{\frac{(l+2)!}{(l-2)!}}\,\tilde a_{B,lm}\,Y_{lm}(\mathbf n).\label{B_iBlm}
\end{eqnarray}

\section{Spherical harmonic transformation and discontinuities}
\label{discrepancy}
\begin{figure}
\centering\includegraphics[scale=.27]{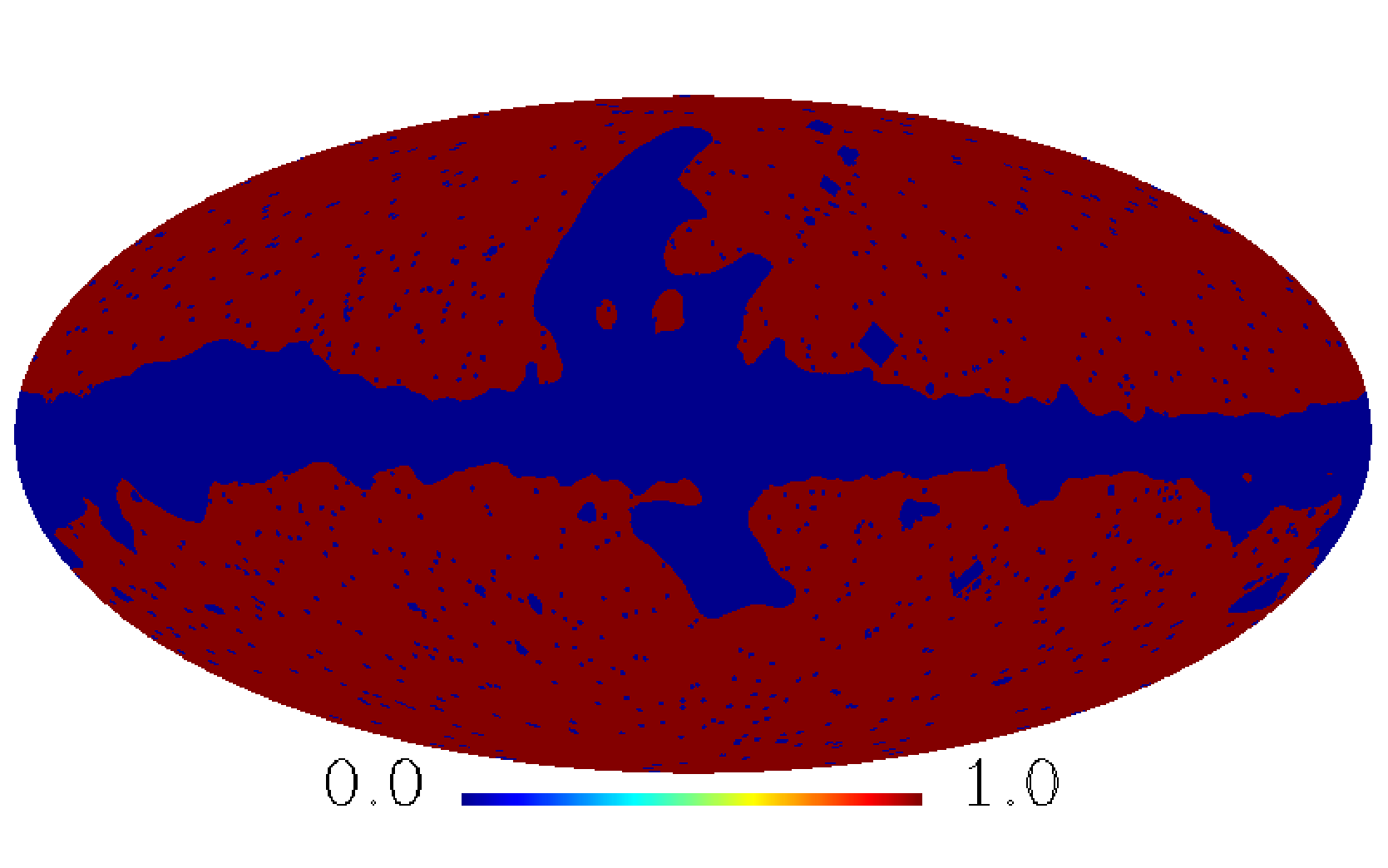}
\centering\includegraphics[scale=.27]{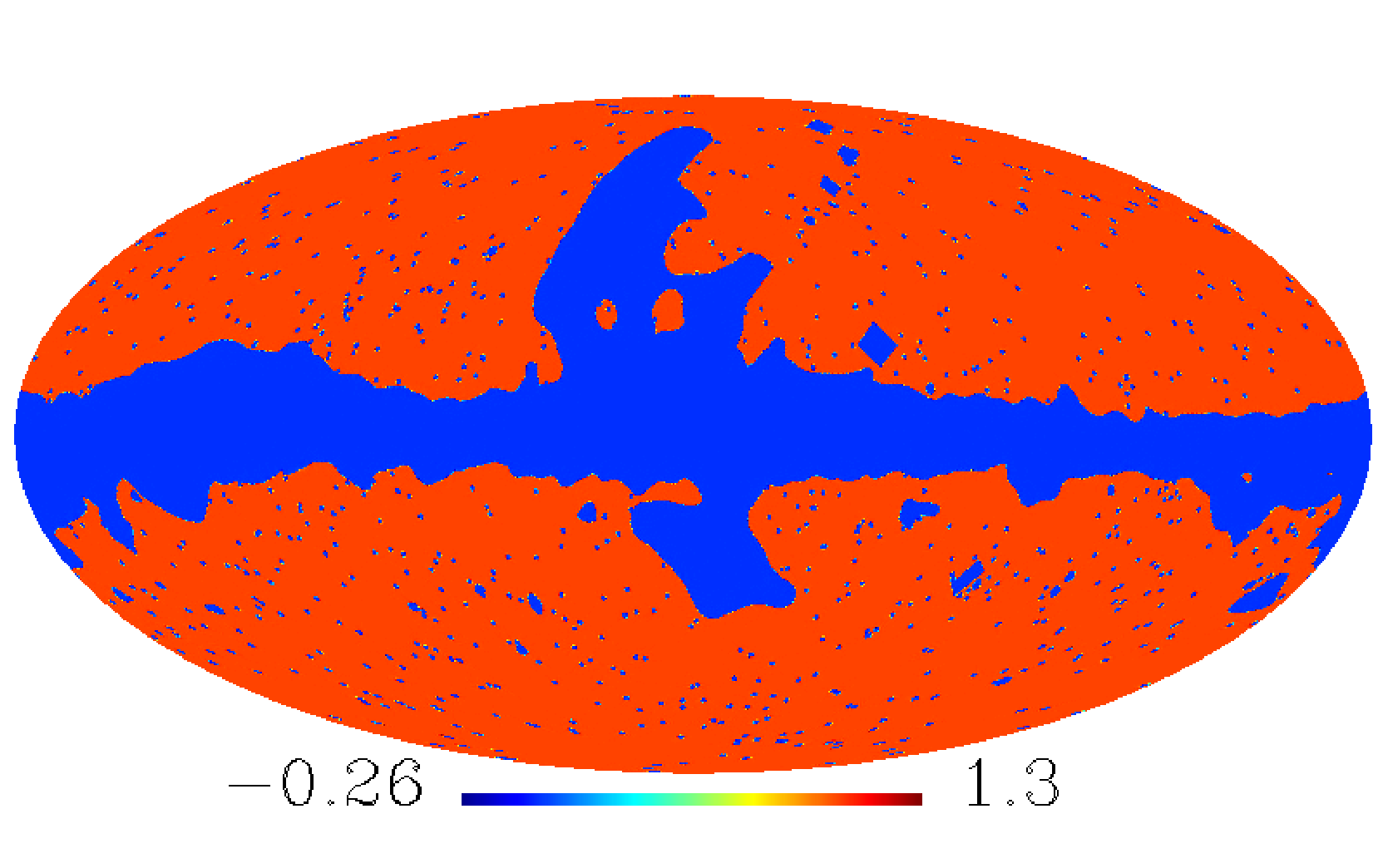}
\caption{Foreground mask: the input (top) and the reconstruction (bottom)}
\label{mask0}
\end{figure}

In deriving Eqs. \ref{E_iElm} and \ref{B_iBlm}, we assumed that the forward and backward spherical transformations retain a masked polarization signal with good accuracy.
\begin{eqnarray}
\tilde Q(\mathbf n)\pm i \tilde U(\mathbf n)\approx \sum^{l_{\mathrm{max}}}_{l,m} \tilde a_{\pm2,lm}\;{}_{\pm2}Y_{lm}(\mathbf n). \label{expansion}
\end{eqnarray}
However, there exists the Gibbs Phenomenon (hearafter GP), which refers to the peculiar manner in which the Fourier representation behaves at a jump discontinuity \citep{Gibbs_Fourier,Gibbs_phenomenon}. 
The Gibbs phenomenon disappear only in the continuum limit (i.e. infinite pixel resolution and summation up to an infinitely high multipole) \citep{Gibbs_Fourier,Gibbs_phenomenon,signal_analysis,medical_imaging,DSP,digital_image_synthesis}. 
Jump discontinuities due to foreground masking lead to the Gibbs phenomenon, making Eq. \ref{expansion} a poor approximation. To investigate the degree of the Gibbs phenomenon, we performed a spherical harmonic transformation of a masked CMB sky map, and reconstructed it from its inverse transformation. Throughout this paper, all spherical harmonic transformation were performed using HEALPix subroutine \texttt{map2alm\_iterative} with three iterations.
To construct the foreground mask, we combined the WMAP team's polarization mask with the point source mask, and prograded it to a higher resolution (HEALPix Nside=2048), which is comparable to the Planck pixel resolution.
The foreground mask described above may be suboptimal for observations other than WMAP (e.g. Planck surveyor).
However, we find it sufficient for our purposes, which are the investigation of E/B leakage caused by cut sky.
For the polarization map, we used the simulated polarization map of $10'$ FWHM, which is described in Section \ref{simulation}.
For spherical harmonic transformation, we set the maximum multipole $l_{\mathrm{max}}$ to 4096, as recommended by HEALPix for a chosen pixel resolution (Nside=2048). 
\begin{figure}[htb!]
\centering\includegraphics[scale=.27]{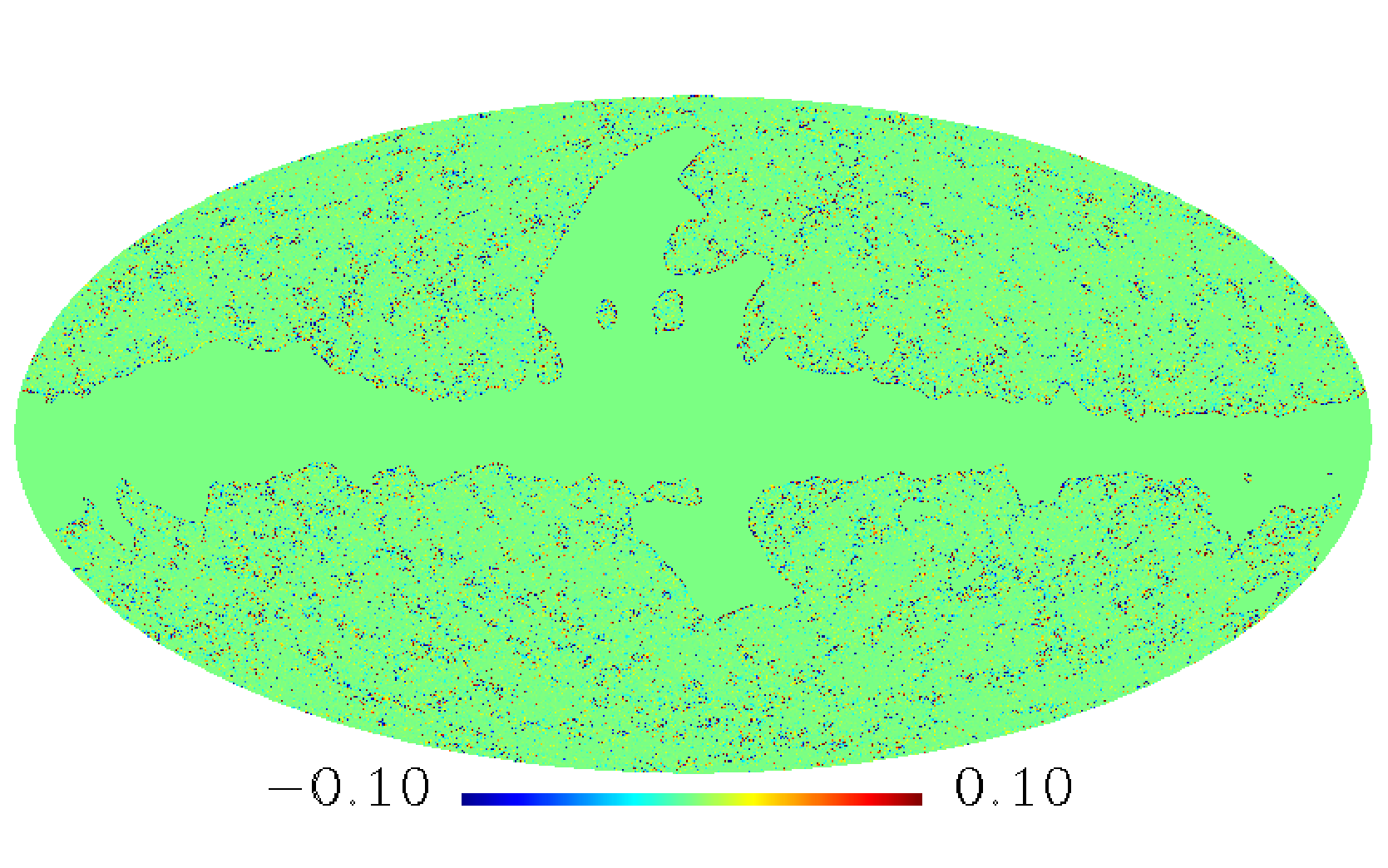}
\centering\includegraphics[scale=.27]{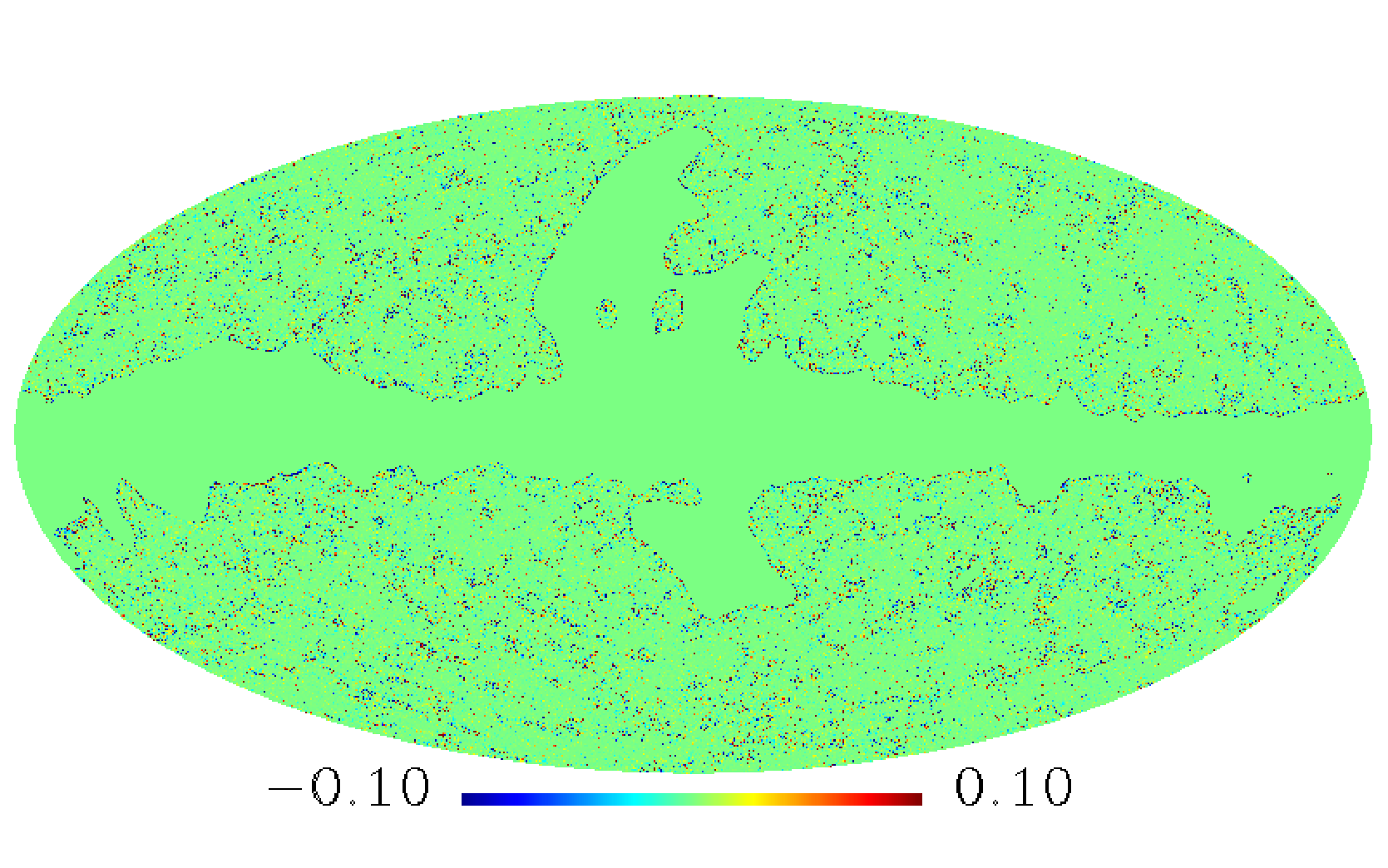}
\caption{Relative difference between the input and its reconstruction: $\Delta Q/Q$ (top) and $\Delta U/U$ (bottom)}
\label{dQU}
\end{figure}
In Fig. \ref{dQU}, we show the relative difference between the input and its reconstruction, respectively, for the Q and U signals,
where we performed a spherical harmonic transformation of the input map and subsequently reconstructed it by the inverse transformation. As shown in Fig. \ref{dQU}, there is a non-negligible discrepancy between the original signal and its reconstruction.
To show the contours of the foreground mask more clearly, we set the minimum and maximum of the pixel values to $-0.1$ and $0.1$, respectively, though the value of the pixels actually lies in much wider ranges $-100 \le \Delta Q/Q\le 158$ and $-160 \le \Delta U/U\le 502$.
For comparison, we performed a spherical harmonic transformation of Q and U maps without the foreground mask and reconstructed Q and U maps by the inverse transformation. We find that the relative difference is found to be at the level of $\lesssim 10^{-4}$. 
The discrepancy is obviously attributed to a foreground mask.
For a further investigation, we performed a spherical harmonic transformation of the foreground mask (HEALPix Nside=2048), and reconstructed the foreground mask by the inverse transformation.
In a manner similar to the polarization map reconstruction, we made the spherical harmonic transformation up to the multipole 4096. In Fig. \ref{mask0}, we show the original mask and its reconstruction.
\begin{figure}[bt!]
\centering\includegraphics[scale=.55]{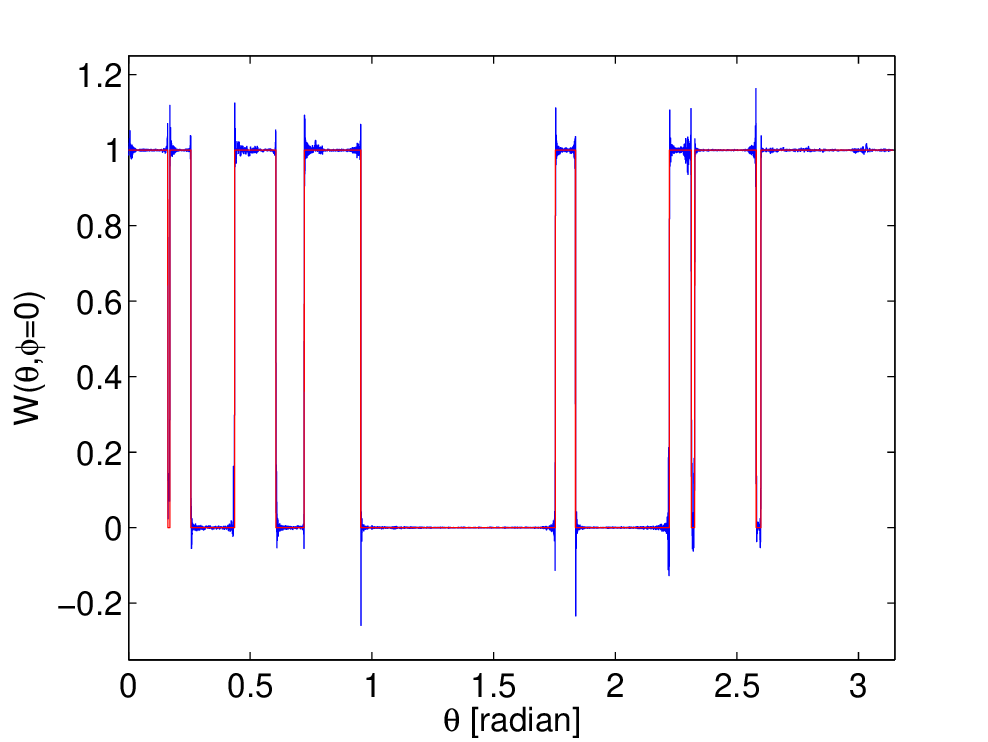}
\caption{$W(\theta, \phi=0)$: the red and blue lines correspond to that of the input and the reconstruction respectively.}
\label{W0}
\end{figure}
As shown in Fig. \ref{mask0}, we find a significant discrepancy between the original mask and its reconstruction.
In Fig. \ref{W0}, we show the values of the mask $W(\mathbf n)$ for a fixed azimuthal angle $\phi=0$ to provide a clearer view, and note
the presence of the Gibbs phenomenon (i.e. the ringing pattern around jump discontinuities). 

\section{Smoothed foreground mask and proper foreground masking}
\label{smoothing}
The Gibbs phenomenon disappears, when a spherical harmonic expansion is made up to an infinitely high multipole.
Hence, the terms of multipoles higher than the truncation point may be associated with the discrepancy shown in Figs. \ref{dQU} and \ref{mask0}.
Accordingly, we may reduce the Gibbs phenomenon by suppressing the terms of multipoles higher than the truncation point.
In image processing, where the Gibbs phenomenon has been known for a long time, Gaussian smoothing has been widely used to mitigate the Gibbs phenomenon \citep{signal_analysis,medical_imaging,DSP,digital_image_synthesis}. To suppress the terms of high multipoles, we consider and use the Gaussian smoothing kernel consistently throughout this work.
\begin{eqnarray}
B(\theta)=\frac{1}{\sqrt{2\pi \sigma^2}}\exp\left(-\frac{\theta^2}{2\sigma^2}\right), \label{B}
\end{eqnarray}
where $\theta$ is a separation angle.
The Gaussian smoothing kernel is a low pass filter, which has the window function \citep{Modern_Cosmology}.
\begin{eqnarray}
B_l=\exp(-l^2 \sigma^2), \label{low_pass}
\end{eqnarray}
where $\sigma=\mathrm{FWHM}/\sqrt{8\ln(2)}$ and FWHM denotes the full width at half maximum of the smoothing kernel.
As discussed previously, we made a spherical harmonic expansion up to $l=4096$, as recommended for the HEALPix Nside=2048 pixel resolution.
We therefore set the FWHM of a smoothing kernel to $15'$, so that the multipoles $l> 4096$ are sufficiently suppressed (e.g. $B_{l=4096}\sim 10^{-25}$).
One may also consider increasing the maximum multipole (i.e. $l_{\mathrm{max}}$) in a spherical harmonic representation.
However, given a fixed pixel resolution, increasing $l_{\mathrm{max}}$ beyond the aliasing limit does not mitigate the Gibbs phenomenon, but instead leads to additional distortion.

\begin{figure}[htb!]
\centering\includegraphics[scale=.55]{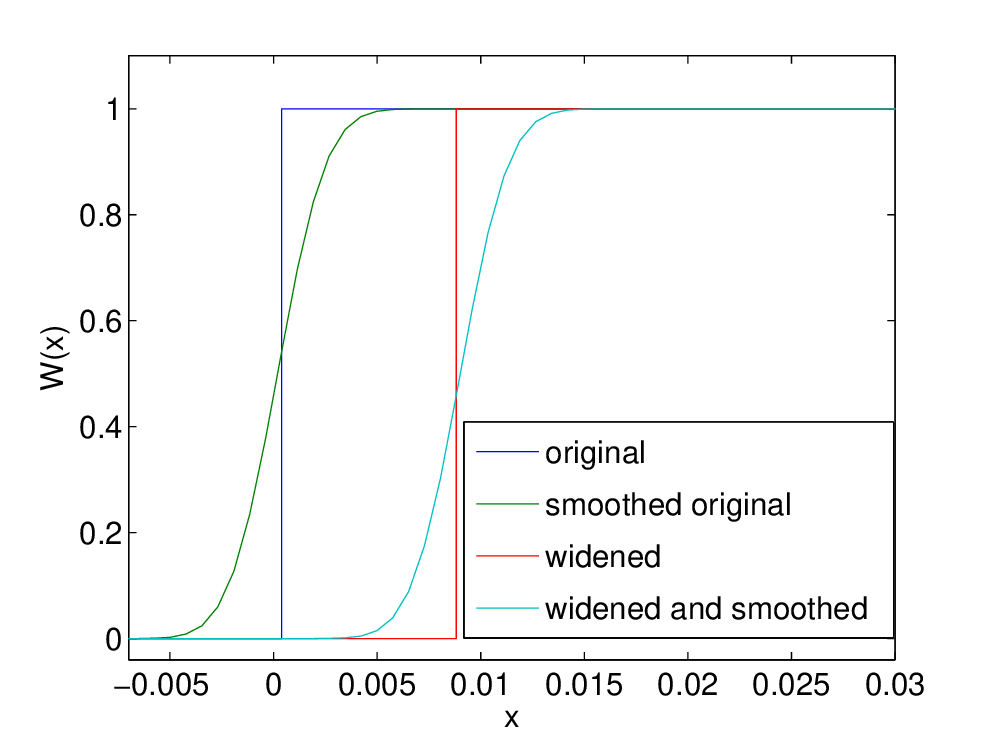}
\caption{$W(x)$: the blue, green, red, and cyan curves correspond to the original mask, the smoothed original mask, the widened mask, and the widened and smoothed mask, respectively.}
\label{w1d}
\end{figure}
\begin{figure}
\centering\includegraphics[scale=.27]{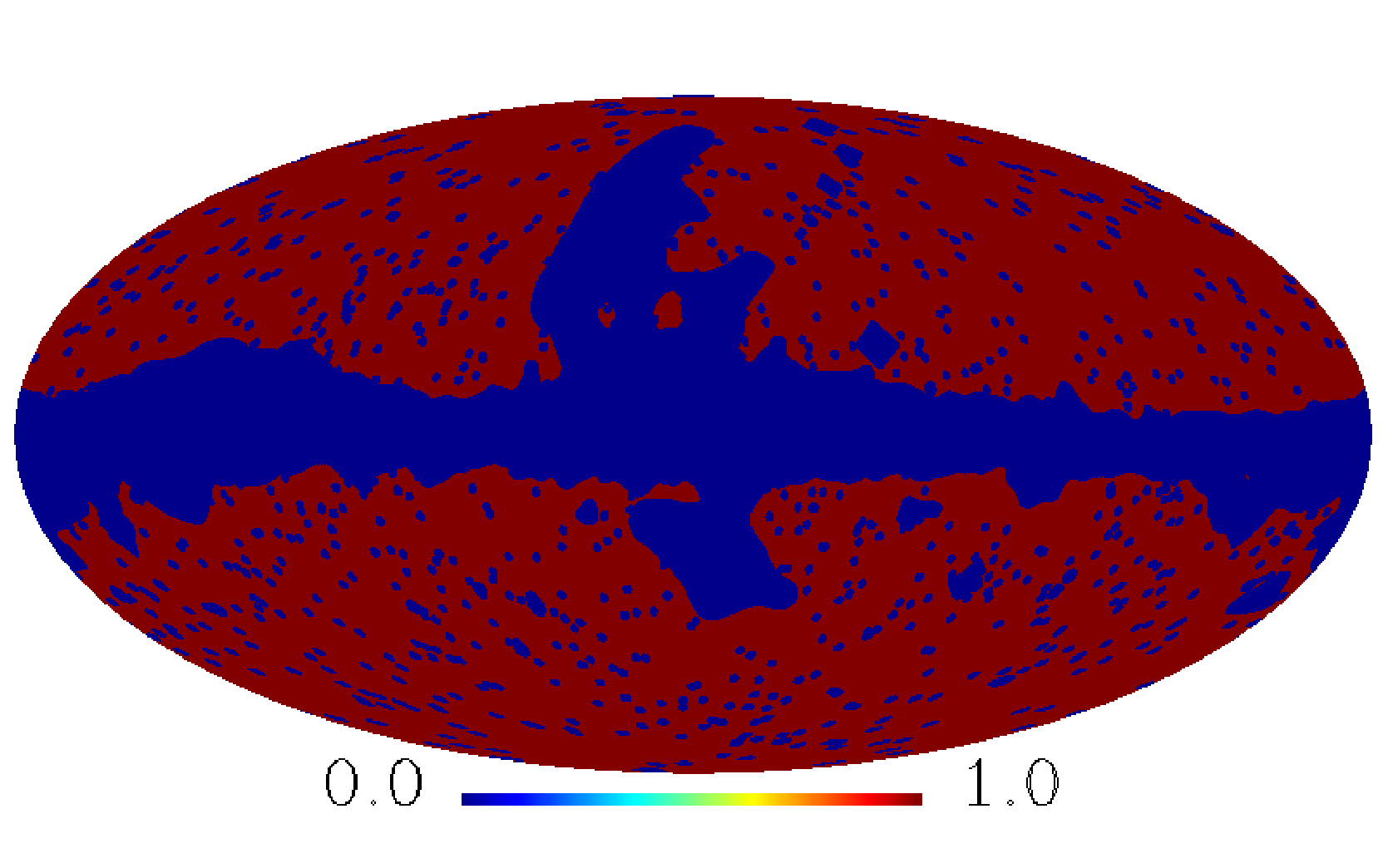}
\centering\includegraphics[scale=.27]{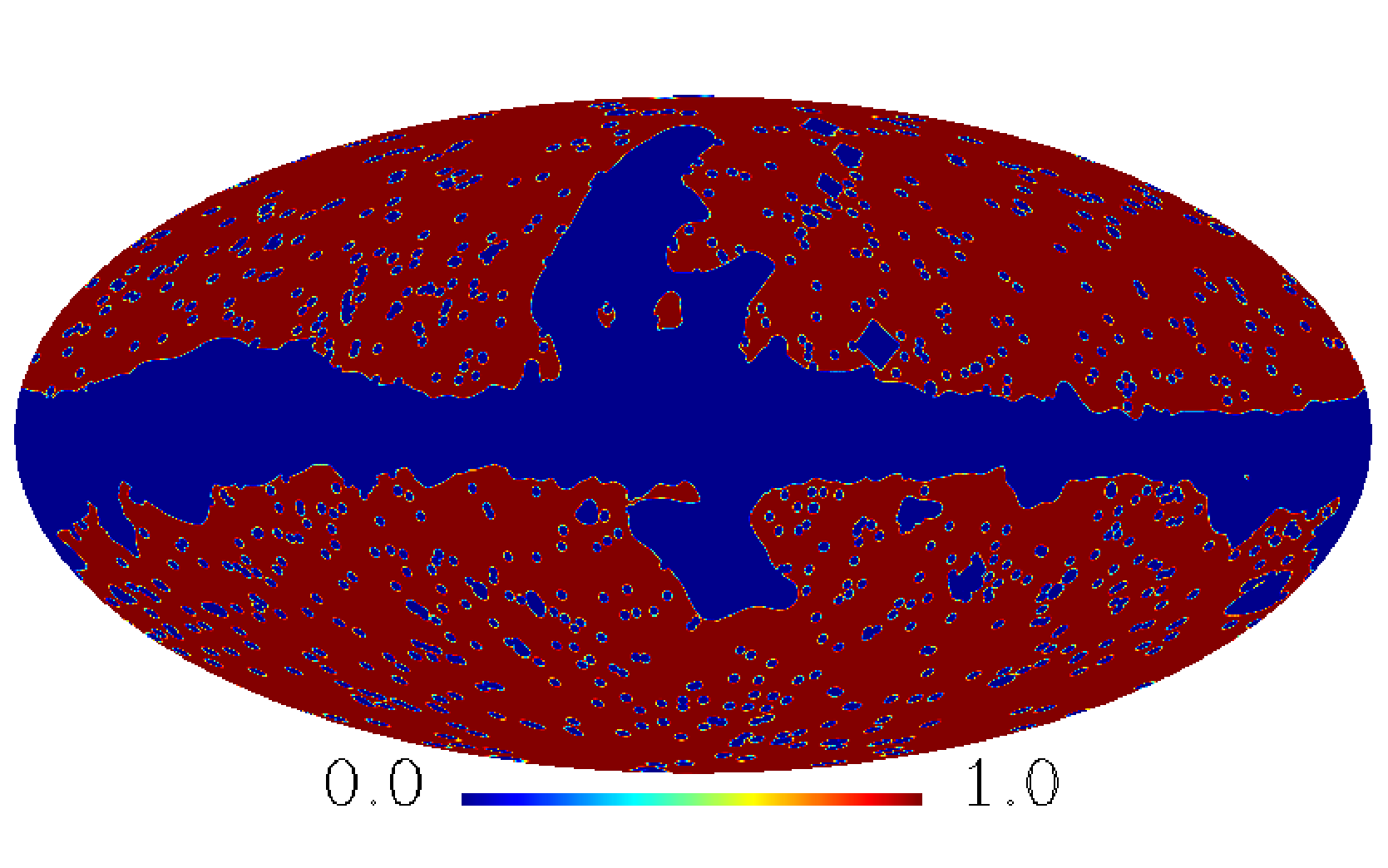}
\caption{Foreground mask: the widened mask (top) and the `processing mask' (bottom)}
\label{mask15}
\end{figure}
Smoothing a foreground mask, we may incur foreground contamination, because pixels, which are zero in the original mask, may be rendered non-zero by the smoothing process. We therefore have to widen the original mask before smoothing so that the originally zeroed pixels may remain zero.
Without loss of generality, we may consider a one dimensional case in a direction normal to the contours of foreground cut.
We assume the original foreground mask and the widened mask to be zero, respectively, for $x\le 0$ and $x\le x_0$, as plotted in Fig. \ref{w1d},
where $x_0$ is a positive number that will be determined later.
We apply smoothing to a widened mask to produce a smoothed mask, which is plotted as a cyan curve in Fig. \ref{w1d}.
Subsequently, we set the pixels ($W(\mathbf n)\le \beta$) to zero, where $\beta$ is a very small number.
In addition to this, we set pixels ($W(\mathbf n)\ge 1-\beta$) to one so that we may maximize the area of constant values to a good approximation. 
From now on, we are going to call this product `a processing mask', and use it for E and B mode decomposition via Eq. \ref{iElm} and \ref{iBlm}.
To decide the value of $\beta$, we investigated the reconstruction accuracy of a smoothed mask, which is mainly limited by the numerical accuracy of the HEALPix package.
Because the associated accuracy is $\sim 10^{-6}$, we set $\beta$ to $10^{-6}$. 
As discussed previously, we wish to ensure that the zero-value pixels of the original mask are zero in the processing mask.
To do that, we wish the cyan curve in Fig. \ref{w1d} to have a value of $\beta$ at $x=0$, so that the pixels $x\le0$ in the `processing mask' may be zero.
The pixel value of a smoothed mask at $x=0$ is given by
\begin{eqnarray}
\beta&=&\int^{\infty}_{x_0} \frac{1}{\sqrt{2\pi \sigma^2}}\exp\left(-\frac{(0-x')^2}{2\sigma^2}\right) dx' \nonumber\\
&=&\frac{1}{2}\left(1-\mathrm{erf}(x_0/\sqrt{2\sigma^2})\right)\label{beta}
\end{eqnarray}
where the integral corresponds to the convolution of the widened mask with the smoothing kernel.
The lower bound of the integral is set to $x_0$, since we smooth a widened mask.
Using Eq. \ref{beta}, we can show that the boundary location $x_0$ of a widened foreground mask is given by
\begin{eqnarray}
x_0=\sqrt{2\sigma^2}\;\mathrm{erf}^{-1}(1-2\beta). \label{x0}
\end{eqnarray}
However, it is computationally prohibitive to solve Eq. \ref{x0}, given an immense number of pixels.
We therefore need to derive a computationally efficient form, for which we may consider smoothing the original mask. In Fig. \ref{w1d}, the smoothed original mask is plotted as a green cyan curve.
The pixel value of the green curve at $x_0$ is then given by
\begin{eqnarray}
\alpha&=&\int^{\infty}_{0} \frac{1}{\sqrt{2\pi \sigma^2}}\exp\left(-\frac{(x_0-x')^2}{2\sigma^2}\right) dx'\nonumber\\
&=&\frac{1}{2}\left(1+\mathrm{erf}(x_0/\sqrt{2\sigma^2})\right).\label{alpha}
\end{eqnarray}
The lower bound to the intergal of Eq. \ref{alpha} and the exponent argument differ from those of Eq. \ref{beta}.
Using Eqs. \ref{x0} and \ref{alpha}, we may easily show that $\alpha=1-\beta$, which corresponds to the value of the cyan curve at $x_0$.
We may therefore produce the widened mask efficiently by smoothing the original mask and setting the pixels ($W(\mathbf n)\le 1-\beta$) to zero.
We emphasize that we have used the same smoothing kernel consistently. In other words, $\sigma$ in Eq. \ref{beta} should be the same as that in Eq. \ref{alpha}.
Implementing the processes described above, we produced a widened mask from the original mask and 
a `processing mask' subsequently from the widened mask.
In Fig. \ref{mask15}, we show the widened mask and the processing mask.
We find that the sky fraction of the `processing mask' corresponds to $0.64$.
Compared to the original mask ($f_{\mathrm{sky}}=0.71$), we find that a decrease in sky fraction is insignificant.
We also confirm that all zero-value pixels of the original mask are zero in the processing mask.

\begin{figure}[htb!]
\centering\includegraphics[scale=.51]{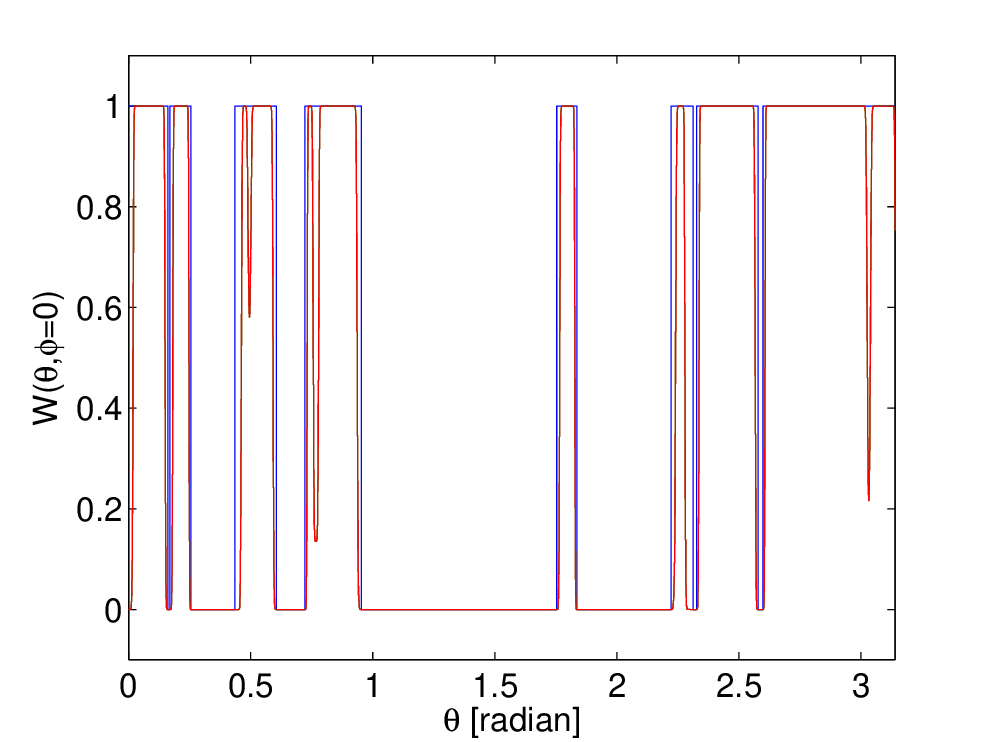}
\caption{$W(\theta, \phi=0)$ of the `processing mask' (a red curve) and its reconstruction: The reconstructed `processing mask' is barely visible, because of the visual indistinguishability between the `processing mask' and its reconstruction. We also show the original mask (a blue curve), which confirms that the zero-value pixels of the original mask are all masked by the `processing mask'.}
\label{W15}
\end{figure}
\begin{figure}[htb!]
\centering\includegraphics[scale=.51]{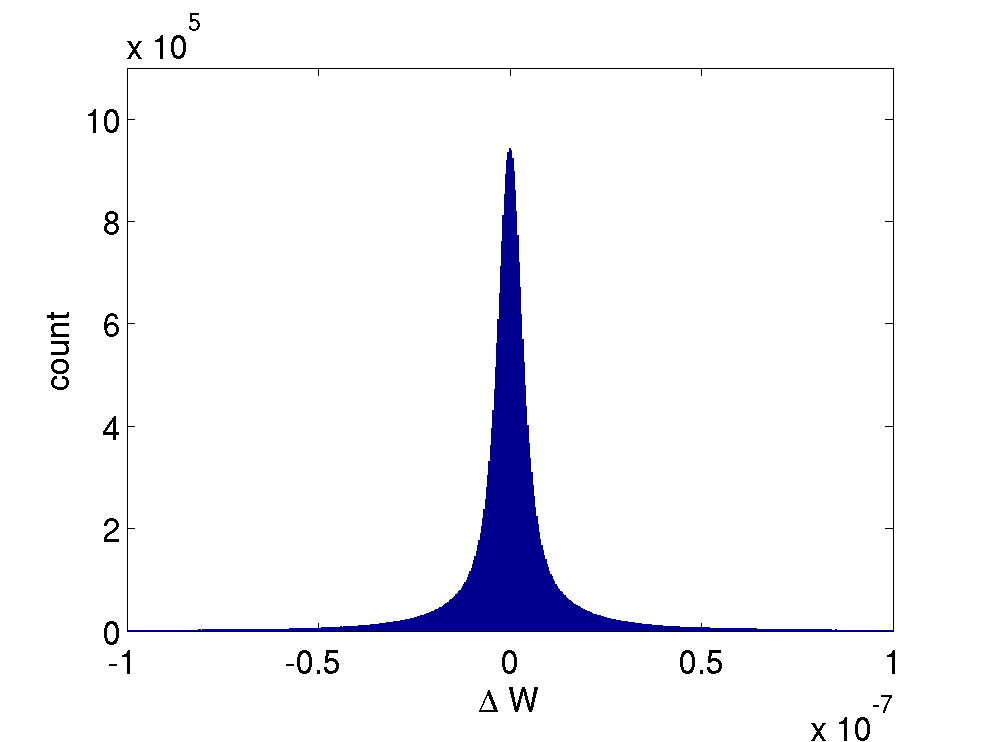}
\caption{Pixel histogram of difference between the `processing mask' and its reconstruction.}
\label{dW15}
\end{figure}
\begin{figure}[htb!]
\centering\includegraphics[scale=.51]{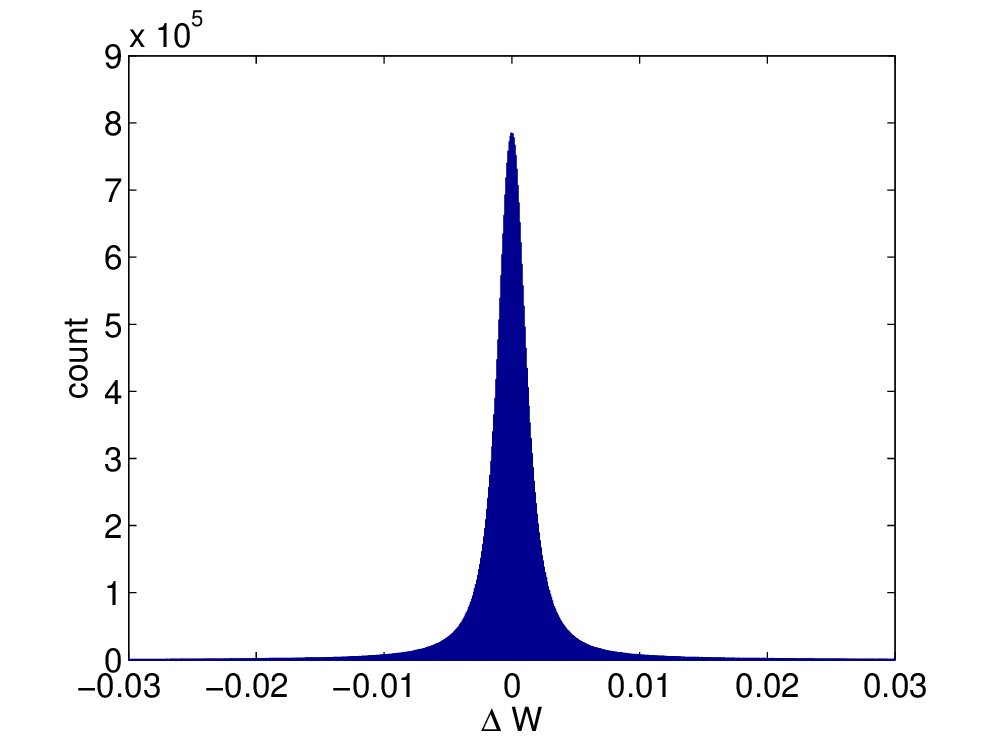}
\caption{Pixel histogram of difference between the `original mask' and its reconstruction.}
\label{dW}
\end{figure}
We performed a spherical harmonic transformation of the `processing mask' and reconstructed it by the inverse transformation.
In Fig. \ref{W15}, we show the `processing' mask as a red curve and its reconstruction for a fixed azimuthal angle $\phi=0$. 
The reconstructed `processing' mask is barely visible, because the `processing mask' and its reconstruction are visually indistinguishable.
In Fig. \ref{dW15}, we show the pixel histogram of difference between the `processing mask' and its reconstruction.
As shown in Figs. \ref{W15} and \ref{dW15}, the reconstruction error is quite negligible, which indicates that the Gibbs phenomenon is effectively suppressed.
For comparison, we show the pixel histogram of the difference between the original mask and its reconstruction in Fig. \ref{dW}.

\section{Application to simulated data}
\label{simulation}
\begin{figure}[htb!]
\centering\includegraphics[scale=.27]{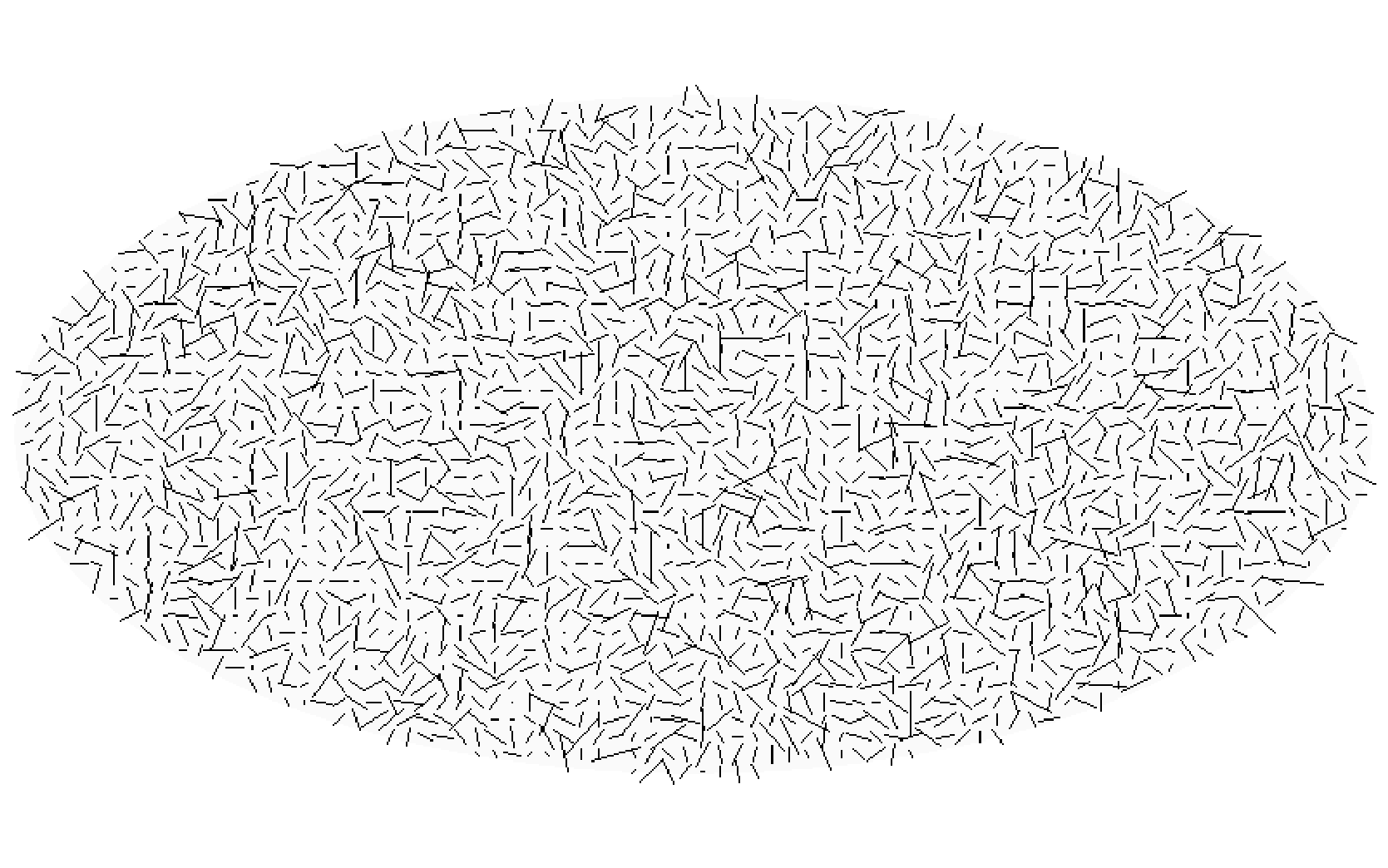}
\caption{Input polarization map (Nside=2048): E mode polarization only.}
\label{input}
\end{figure}

\begin{figure}
\centering\includegraphics[scale=.27]{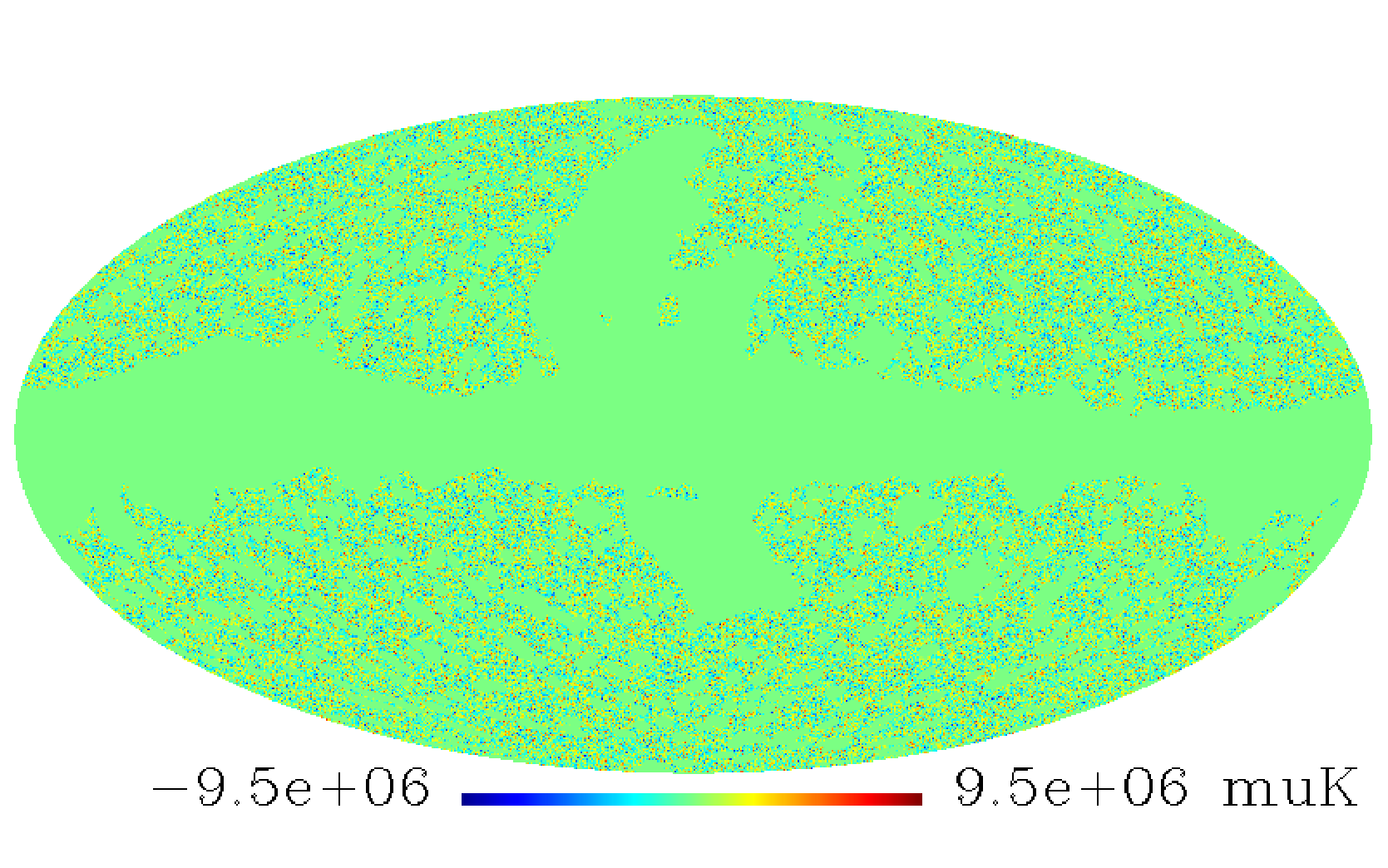}
\centering\includegraphics[scale=.27]{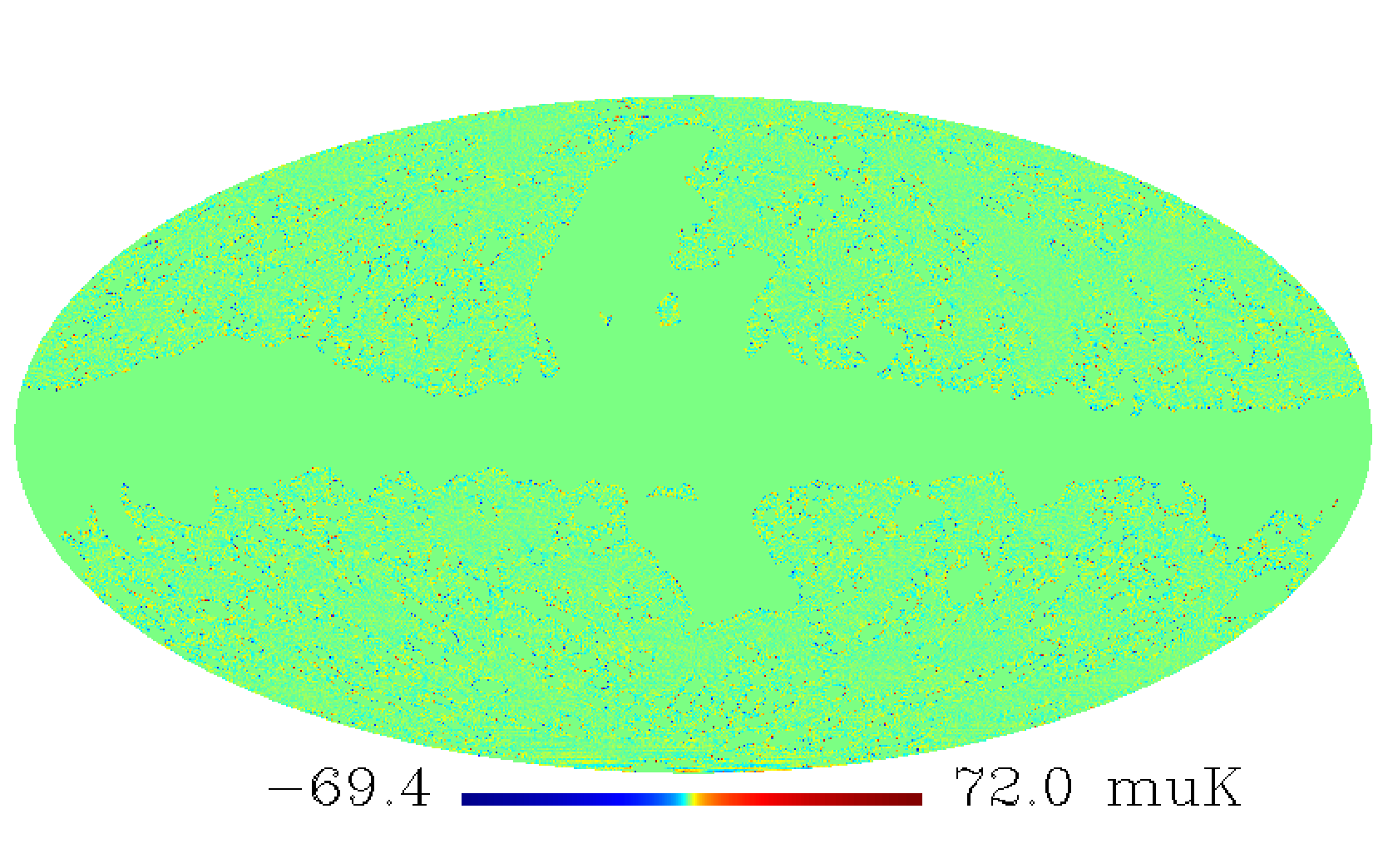}
\caption{$\tilde E$ (top), $\tilde B$ map (bottom)}
\label{output1}
\end{figure}

Using the WMAP concordance $\Lambda$CDM model, we produced simulated Stokes parameter Q and U over the whole sky with a HEALPix pixel resolution Nside=2048.
For the beam  of the observation, we assumed a $10'$ FWHM, which corresponds to Planck HFI beam at 100 GHz channel. 
In Fig. \ref{input}, we show our simulated polarization map, where the orientation and length of headless arrows indicates polarization angle and amplitude, respectively. 
We simulated the inputmap of no B mode polarization so that any non-zero values in a decomposed B map may be attributed to leakage.
For a foreground mask, we use a `processing mask' as shown in Fig. \ref{mask15}.
We applied the `processing mask' to the simulated map, and produced $\tilde E$ and $\tilde B$ maps via Eqs. \ref{iElm}, \ref{iBlm}, \ref{ialm_E}, and \ref{ialm_B}. 
In Fig. \ref{output1}, we show the $\tilde E$ and $\tilde B$ maps, where the unmasked pixels correspond to a sky fraction $f_{\mathrm{sky}}=0.64$.
The magnitude of the E and B maps shown in Fig. \ref{output1} are much higher than the input polarization map, because of the prefactor $\sqrt{(l+2)!/(l-2)!}$ in the definitions given by Eqs. \ref{Elm} and \ref{Blm}.

\begin{figure}[htb!]
\centering\includegraphics[scale=.55]{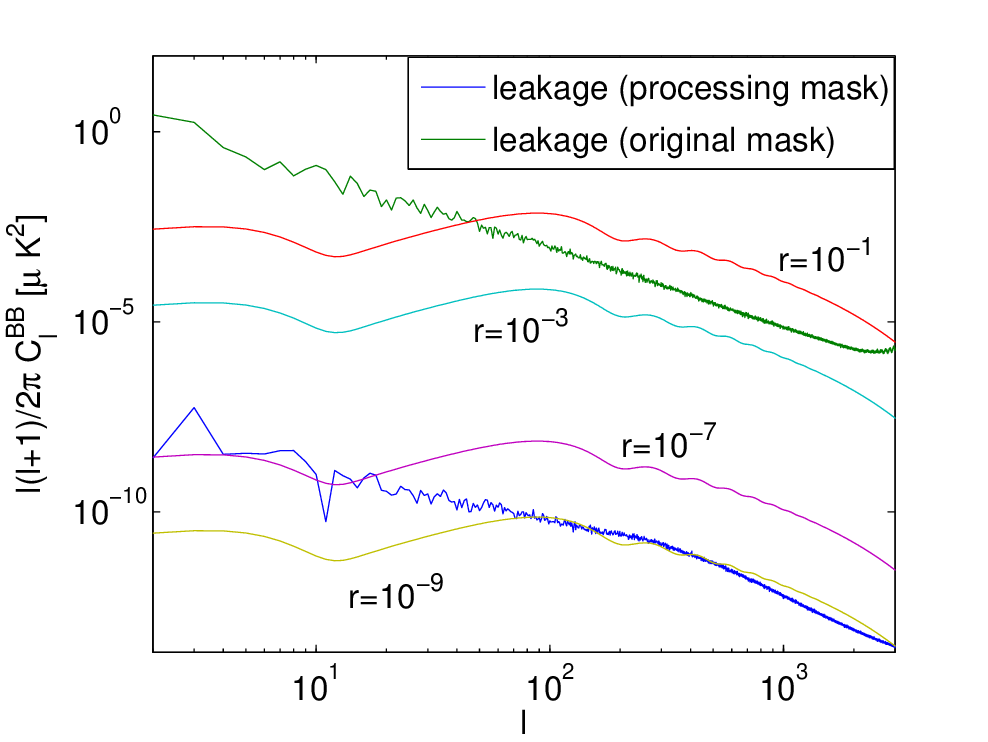}
\caption{Leakage power spectrum and primordial B mode power of various tensor-to-scalar ratio $r$: 
blue and green curves denote the leakage power produced, respectively, by a `processing mask' and an original mask.}
\label{leakage}
\end{figure}

\begin{figure}
\centering\includegraphics[scale=.55]{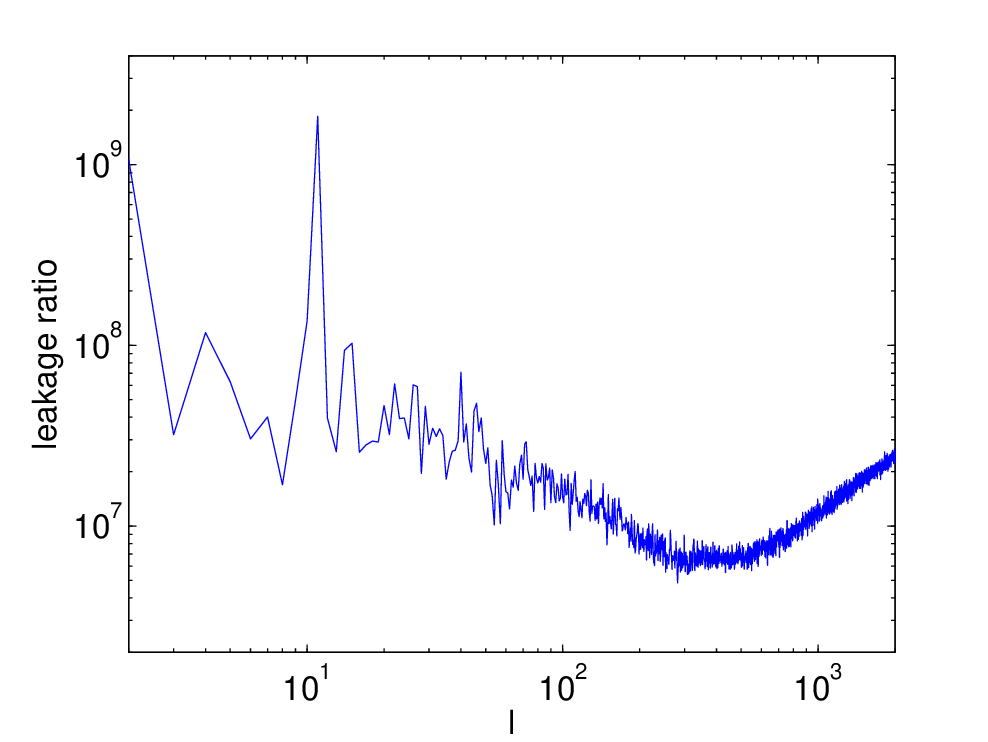}
\caption{Ratio of the leakage power produced by the original mask to that produced by the `processing mask'}
\label{ratio}
\end{figure}

From unmasked pixels of the $\tilde B$ map, we estimated the leakage power spectrum using the pseudo $C_l$ method \citep{pseudo_Cl,MASTER}.
The power spectrum is usually estimated using the pseudo $C_l$ method at high multipoles ($l> 30$), while other methods are adopted at the low multipoles ($l\le 30$) \citep{Bond:likelihood,Gibbs_power,WMAP3:temperature,hybrid_estimation}.
However, we find that using the pseudo $C_l$ method is good enough for leakage power estimation.
In Fig. \ref{leakage}, we show the leakage power spectrum (blue curve) and the B mode power spectrum of various values of the tensor-to-scalar ratio $r$. 
As shown in Fig. \ref{leakage}, we find that the leakage power (blue curve) is comparable to or smaller than B mode power spectrum of tensor-to-scalar ratio $r\sim 1\times10^{-7}$.
For comparison, we repeated the analysis, using the original mask (the top figure in Fig. \ref{mask0}).
In Fig. \ref{leakage} and \ref{ratio}, we plotted the power spectrum of the leakage produced by the original mask as a green curve, and the ratio of the blue curve to the green curve. As shown in Figs. \ref{leakage} and \ref{ratio}, we reduced the leakage power by a factor between $10^{6}$ and $10^{9}$, while losing only a sky fraction $0.07$.
We also confirm that all the zero-value pixels in the original mask are blocked by our `processing mask'.


\section{Discussion}
\label{Discussion}
We have investigated the E and B mode decomposition of masked CMB sky data. 
In real space, the E and B mode decomposition operators involve only differentials of CMB polarization.
Therefore, we may, in principle, construct E and B maps in real space from incomplete sky data without E/B confusion. 
However, it is impractical to apply second derivatives to observation data. Therefore, we usually resort to a spherical harmonic transformation and its inverse transformation when performing a E/B decomposition. In spherical harmonic representation, jump discontinuities in a cut sky produce the Gibbs phenomenon, unless spherical harmonic expansion consists of infinitely high multipoles. Fortunately, we may suppress the Gibbs phenomenon by smoothing a foreground mask.
This smoothing approach is similar to the apodized foreground mask discussed in other works \citep{Smith:pseudo_EB,edge_taper,EB_pixel,EBpixel_Zhao}.
By smoothing a foreground mask, we incur foreground contamination, because zero-value pixels in the original mask may be rendered non-zero by the smoothing process.
We have therefore investigated how to derive an optimal foreground mask, which ensures proper foreground masking without any unnecessary loss of sky fraction and suppresses the Gibbs phenomenon at the same time.

We have applied our method to a simulated map of the pixel resolution comparable to the Planck satellite. The simulation shows that the leakage power in unmasked pixels ($f_{\mathrm{sky}}=0.64$) is comparable to or smaller than the unlensed CMB B mode power spectrum of tensor-to-scalar ratio $r\sim 1\times10^{-7}$. For comparison, we repeated the analysis using the original mask, and found that we reduced the leakage power by a factor of $10^{6} \sim 10^{9}$ with the loss of only a sky fraction $0.07$. 
We also confirm that all the zero-value pixels of the original mask are blocked by our processing mask.

Once we construct pseudo E and B maps using the discussed method, we will be able to perform analyses such as a power spectrum estimation or a Gaussianity study, without worrying about the statistical confusion between the E and B mode. The application of our work to the Planck data will surely enhance the detectability of primordial tensor perturbation.

\begin{acknowledgements}
We thank Pavel Naselsky for useful discussion. We thank the anonymous referee for helpful comments.
We acknowledge the use of the HEALPix package \citep{HEALPix:Primer,HEALPix:framework}.
This work is supported by FNU grant 272-06-0417, 272-07-0528 and 21-04-0355. 
This work is supported in part by Danmarks Grundforskningsfond, which allowed the establishment of the Danish Discovery Center.
\end{acknowledgements}

\bibliographystyle{aa}
\bibliography{/afs/mpa/home/kim/Documents/bibliography}
\end{document}